\documentclass[aps,twocolumn,prl]{revtex4-2}

\usepackage{graphicx,amsmath, amsfonts,bm, color}
\usepackage{float}
\usepackage{xcolor,hyperref}
\hypersetup{
   colorlinks,
   linkcolor={blue!50!black},%{red!80!black},
   citecolor={blue!50!black},
   urlcolor={blue!80!black}
}
\usepackage{enumitem}

\renewcommand{\=}{\!=\!}

\newcommand{\sigmac}{\sigma_\mathrm{c}}
\newcommand{\deltac}{\delta_\mathrm{c}}
\newcommand{\Cw}{C_\mathrm{w}}
\newcommand{\fw}{f_\mathrm{w}}
\newcommand{\dfw}{f'_\mathrm{w}}
\newcommand{\kw}{\vert k \vert \omega}

\newcommand{\Gc}{G_\mathrm{c}}

\newcommand{\Cr}{c_\mathrm{r}}
\newcommand{\Cf}{c_\mathrm{f}}
\newcommand{\Cs}{c_\mathrm{s}}
\newcommand{\Cd}{c_\mathrm{d}}
\newcommand{\Vc}{v_\mathrm{c}}
\newcommand{\wv}{\omega_v}
\newcommand{\wo}{\omega_0}

\begin{document}

\title{Dynamic crack front deformations in cohesive materials}
\author{Thibault Roch$^{1}$}
\author{Mathias Lebihain$^{2}$}
\author{Jean-Fran\c{c}ois Molinari$^{1}$}
\thanks{jean-francois.molinari@epfl.ch}
\affiliation{$^{1}$Civil Engineering Institute, Materials Science and Engineering Institute, Ecole Polytechnique F\'ed\'erale de Lausanne, Station 18, CH-1015 Lausanne, Switzerland\\
$^{2}$Laboratoire Navier, CNRS (UMR 8205), {\'E}cole des Ponts ParisTech, Universit{\'e} Gustave Eiffel, 6-8 avenue Blaise Pascal, 77455 Marne-la-Vall{\'e}e, France}

\date{\today}

%TC:break Abstract
\begin{abstract}

   Crack fronts deform due to heterogeneities, and inspecting these deformations can reveal local variations of material properties, and help predict out of plane damage. Current models neglect the influence of a finite dissipation length-scale behind the  crack tip, called the process zone size. The latter introduces scale effects in the deformation of the crack front, that are mitigated by the dynamics of the crack. We provide a theoretical framework for dynamic crack front deformations in heterogeneous cohesive materials and validate its predictions using numerical simulations.

\end{abstract}

\maketitle

\paragraph*{Introduction}

The propagation of fronts, defining the border between two distinct phases, occurs in numerous physical context such as paper wetting \cite{balankin_kinetic_2006}, combustion \cite{maunuksela_kinetic_1997}, polymerization \cite{lloyd_spontaneous_2021} and fracture mechanics \cite{schmittbuhl_direct_1997}. Fronts usually roughen due to interaction with heterogeneities. In fracture mechanics, a front marks the spatial separation between intact material and crack, and is thereby called a crack front. It deforms as a consequence of the heterogeneous landscape of toughness, the material resistance to crack propagation. Understanding how these deformations occur allow rationalizing the properties of composite materials \cite{lazarus_perturbation_2011,bonamy_failure_2011}. In addition, the transition between faceting and micro-branching for fast crack propagation is thought to be related to high in-plane curvature of the front \cite{kolvin_nonlinear_2017}. Studying the dynamics of front deformations is thus key to unraveling the complex dynamics of heterogeneous dynamic rupture. Coplanar crack propagation is usually studied using perturbative approaches, such as the first-order model derived by Rice \cite{rice_first-order_1985} based on the weight functions theory of Bueckner \cite{bueckner_weight_1987}. This approach has then been extended to dynamic rupture \cite{willis_dynamic_1995,movchan_dynamic_1995} and also to higher orders \cite{leblond_second-order_2012,vasoya_geometrically_2013,kolvin_nonlinear_2017}. This framework has been successfully applied to the deformation of crack front for various shapes of defects \cite{chopin_crack_2011,vasoya_geometrically_2013,xia_toughening_2012} as well as predicting the effective toughness of heterogeneous materials \cite{patinet_quantitative_2013,xia_adhesion_2015,lebihain_towards_2021} and rationalizing the intermittent dynamics of crack front propagation in disordered media \cite{bares_aftershock_2018}. These models are however built on the linear elastic fracture mechanic (LEFM) framework and thereby assume that the dissipation at the crack tip occurs in a finite region, the \emph{process zone}, of negligible size. As a consequence, LEFM based models are bound to treat each asperity scale indifferently. Yet, elasticity is expected to break down along a finite region at the tip of the crack and heterogeneities smaller or larger than this length-scale are expected to affect the crack dynamics differently \cite{barras_interplay_2017,kammer_length_2016}. Cohesive zone models of fracture \cite{dugdale_yielding_1960,barenblatt_processzone_1962} allow considering a finite dissipation length-scale through the introduction of stresses resisting the crack opening near the tip over a finite length, the process zone size. Regarding crack distortion, a recent theoretical study \cite{lebihain_quasi-static_2022} shed light on the importance of considering the process zone size for quasi-static cracks. The presence of a finite dissipation length-scale (i) controls the stability of crack fronts and (ii) introduces scale effects in the pinning of crack fronts by heterogeneities of fracture energy, and these effects are strongly dependent on how the toughness variations are achieved. For dynamic rupture, the process zone size is known to shrink with increasing propagation velocity, thus increasing the importance of this length-scale relatively to the size of the heterogeneities \cite{rice_mechanics_1980,morrissey_crack_1998,svetlizky_classical_2014}. In this manuscript, we first investigate numerically the dynamic crack front deformations of co-planar cracks loaded under normal tensile stress (mode-I) conditions and propagating through a heterogeneous toughness field. We solve this problem using our open-source implementation \cite{roch_cracklet_2022} of the spectral boundary integral formulation of the electrodynamics equations \cite{Geubelle1995,Breitenfeld1998} and study the influence of toughness heterogeneities arising from heterogeneities of i) peak strength and ii) process zone size. We then extend the theoretical model of \cite{lebihain_quasi-static_2022} to dynamic rupture and compare the numerical results with the newly derived \emph{ dynamic cohesive line tension model} for a broad range of parameters, thus providing a validation of this model. All in all, we provide a comprehensive framework describing dynamic crack front deformations for cohesive materials.

%%%%%%%%%%%%%% Figure %%%%%%%%%%%%%%%%%%%
\begin{figure*}[ht!]
\centering
\includegraphics[width=1\textwidth]{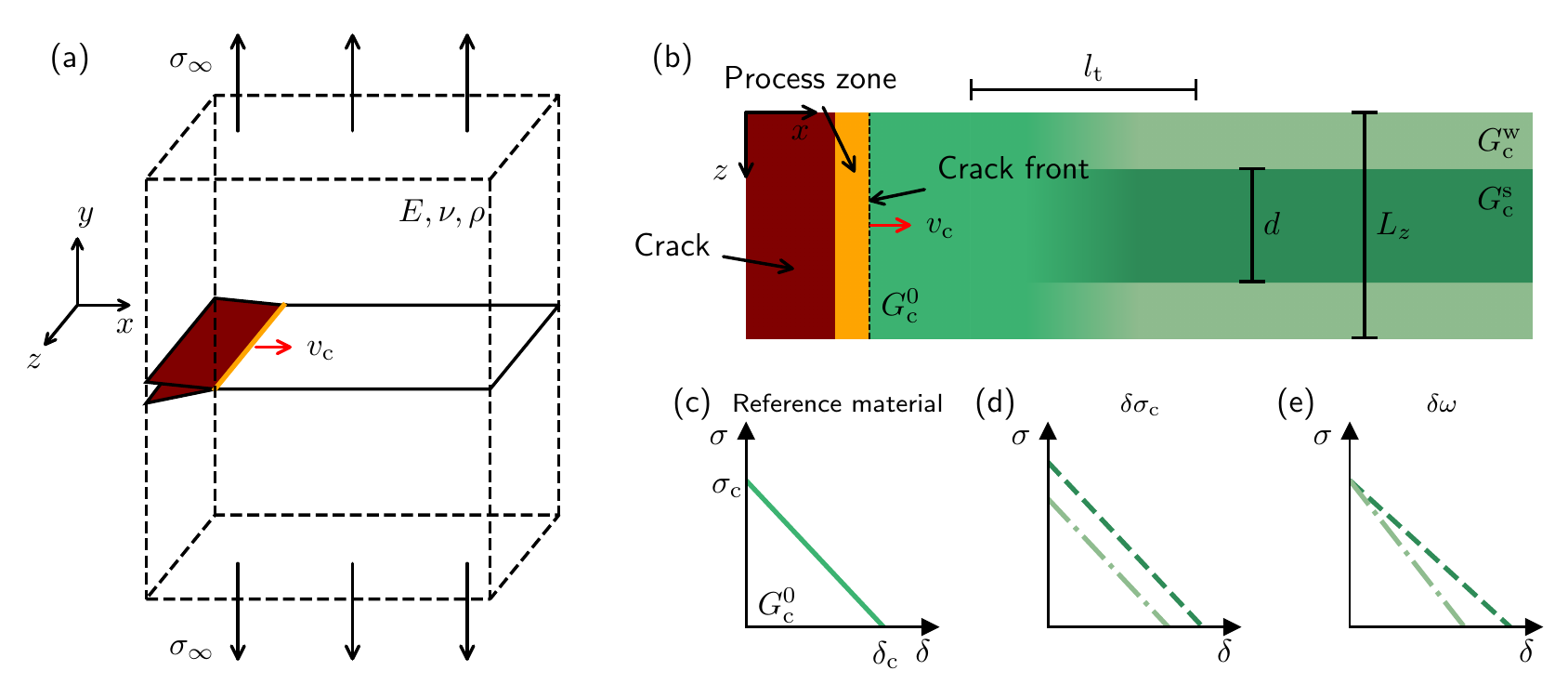}
\caption{(a) Two identical semi-infinite elastic bodies of section $L_x,L_z$ are in contact at a planar interface located at $y=0$. Periodic boundary conditions are imposed for the $x$ and $z$ axes. The bodies are loaded under normal tensile stress that drives a crack through a heterogeneous toughness field in the positive $x$ direction at a constant velocity $\Vc$. (b) The layout of the interface: the crack is in brown, the finite process zone in orange, and the toughness field is represented by shades of green. The crack front is the separation between the process zone and the intact material and is shown with the dashed black line. The toughness is slowly varied from its reference value $\Gc^0$ to respectively $\Gc^{\mathrm{w}}$ and $\Gc^{\mathrm{s}}$ along the transition length $l_t$. (c) Traction-separation law for the reference material. The contrast in toughness can be achieved by (d) changing the peak strength $\sigmac$ but keeping the process zone size equal, or (e) changing the quasi-static process zone $\wo$ size but keeping the peak strength constant.}
\label{fig:fig1}
\end{figure*}
%%%%%%%%%%%%%%%%%%%%%%%%%%%%%%%%%%%%%%%%%%%%%%%

\paragraph*{Problem description}

We consider two semi-infinite elastic bodies of section $L_{x},L_{z}$ that are in contact along a planar interface located at $y=0$ (see Fig.~\ref{fig:fig1}a). Periodic boundary conditions are imposed in the $x$ and $z$ directions. The bodies are loaded under mode-I condition that drives a cohesive crack through a planar interface (crack in brown, process zone in orange in Fig.~\ref{fig:fig1}b) in the positive $x$ direction at a constant velocity $\Vc$. The propagation in the $-x$ direction is prevented. The crack initially propagates inside a homogeneous field of reference toughness $\Gc^0$. The interface properties are then gradually changed along a distance $l_t$ towards an $x$ invariant field composed of a stripe of larger toughness $\Gc^{\mathrm{s}}$ (dark green) of width $d$ embedded in a weaker toughness field $\Gc^{\mathrm{w}}$ (light green). The average toughness in the $z$ direction is kept equal to the reference one, $(\Gc^{\mathrm{s}} + \Gc^{\mathrm{w}} )/ 2 = \Gc^0$, resulting in an effective toughness in the weak pinning regime that is equal to $\Gc^0$ \cite{lebihain_quasi-static_2022}. The gradual transition of properties allows reducing the oscillations of the crack front deformations, see \cite{supplemental} for more details. In this manuscript, we use $d = L_z /2$, $L_x = 8 L_z$. We study the propagation for only $x<0.75L_x$ to neglect the effect of periodic boundary conditions. We employ a linear cohesive law (see Fig.~\ref{fig:fig1}c) to describe the cohesive behavior of the interface, for which the stress decays linearly from a peak value $\sigmac$ to $0$ with the opening $\delta$ up to a critical value $\deltac$

\begin{equation}
    \label{eq:CohesiveLaw}
    \sigma^{str}(x,z,t) = \sigmac(x,z) \max\left[  1-\delta(x,z,t)/ \deltac(x,z),0 \right]
\end{equation}

For the linear slip weakening law, the process zone size at rest $\wo$ can be estimated as $\wo \simeq 0.731 (1-\nu) \mu \deltac / \sigmac$ \cite{viesca_numerical_2018}, with $\nu$ and $\mu$ respectively the Poisson's ratio and the shear modulus of the bulk.

%%%%%%%%%%%%%% Figure %%%%%%%%%%%%%%%%%%%
\begin{figure}[ht!]
%\centering
\includegraphics[width=0.5\textwidth]{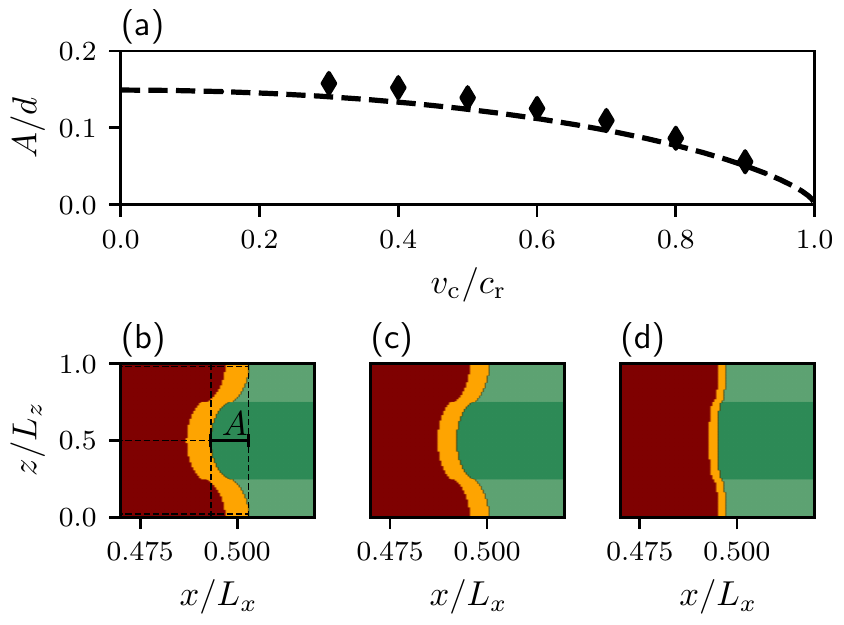}
\caption{(a) Scaling of the amplitude $A / d$ of the front deformations with the propagation velocity $\Vc / \Cr$. $A$ is defined as the distance along $x$ between the most advanced points in the process zone size at $z = 0$ and at $z = 0.5 L_{z}$ as shown in panel (b). The dashed black line corresponds to the prediction of the classical line tension model $A_{\mathrm{lefm}} / d$ corrected to take into account the dynamic stiffening of the front by the term $D_I(\Vc)$, see details in the text. (b)-(c)-(d) : snapshots of the crack front deformation for, respectively $\Vc / \Cr = 0.3 , 0.6 , 0.9$. Note that the $x$-scale and $z$-scale are different. The crack is in brown, the process zone in orange, the strong toughness in dark green and the weak one in light green. The toughness contrast in these simulations is $\Delta \Gc / \Gc^0  = 0.4$.}
\label{fig:fig2}
\end{figure}
%%%%%%%%%%%%%%%%%%%%%%%%%%%%%%%%%%%%%%%%%%%%%%%

%%%%%%%%%%%%%% Figure %%%%%%%%%%%%%%%%%%%
\begin{figure*}[ht!]
%\centering
\includegraphics[width=1\textwidth]{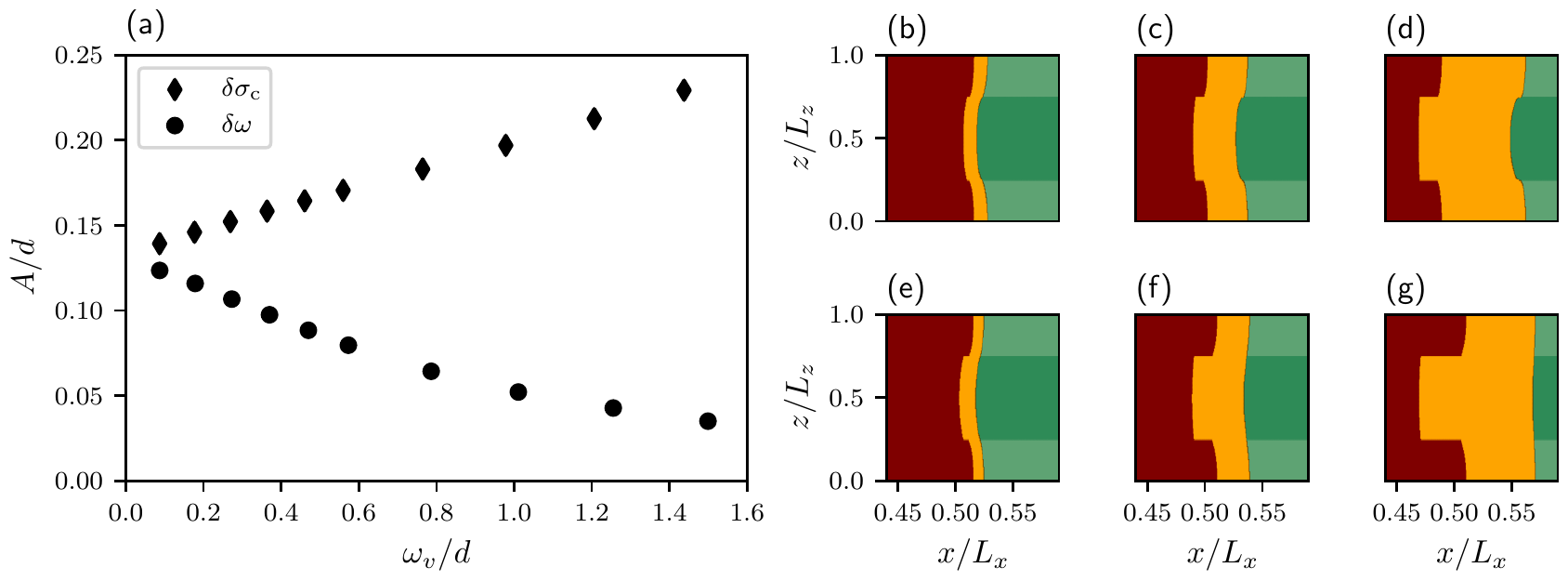}
\caption{(a) Scaling of the amplitude $A$ of the front deformations with the process zone size $\wv$ for heterogeneities of constant process zone size (diamonds, snapshot shown in (b)-(c)-(d) for $\wv /d \sim 0.2 , 0.6, 1.25$) and constant peak strength (circles, snapshots shown in (e)-(f)-(g) for $\wv /d \sim 0.2 , 0.6, 1.25$). For the latter, the value of $\wv$ is the average of $\wv(z)$ over the crack front. The crack velocity in these simulations is $\Vc = 0.5 \Cr$.}
\label{fig:fig3}
\end{figure*}
%%%%%%%%%%%%%%%%%%%%%%%%%%%%%%%%%%%%%%%%%%%%%%%

The opening is defined as the difference between the displacement fields of the top and bottom solids. In this work, we investigate two types of heterogeneities: (1) heterogeneities of peak strength $\sigmac$ with equal process zone size (see Fig.~\ref{fig:fig1}d) or (2) heterogeneities of varying quasi-static process zone size $\wo$ with constant peak strength (see Fig.~\ref{fig:fig1}e). The toughness contrast is defined as $\Delta \Gc = \Gc^{\mathrm{s}} - \Gc^{\mathrm{w}}$. The problem is solved  by conducting full-field dynamic calculations, using an in-house open-source implementation of the spectral boundary integral method \cite{Geubelle1995,Morrissey1997,Breitenfeld1998} called cRacklet \cite{roch_cracklet_2022}. This method relates the displacements $\bm{u}^{\pm}$ of the fracture plane to the stresses $\bm{\tau}$ acting on it. The details of the method are available in \cite{supplemental}. During a typical simulation, the crack front is initially perfectly straight. It starts deforming when it reaches the heterogeneous field of toughness. The dynamic deformation of the crack front is mediated by the propagation of crack front waves \cite{morrissey_crack_1998,fekak_crack_2020,dubois_dynamic_2021}, resulting in the front oscillating over an equilibrium configuration, see \cite{supplemental} for details. We measure the amplitude $A$ of the front deformation as the distance between the most advanced point in the process zone at the axis of the strong band and at the axis of the weak band, as shown in Fig.~\ref{fig:fig2}b. Preliminary to the study of dynamic and process zone effects on the crack deformations, we verified that our numerical model accurately results in a linear increase of front deformation amplitude with the toughness contrast for a given velocity, see \cite{supplemental}.

\paragraph*{Crack propagation velocity}

First, we investigate the effect of the propagation velocity on the dynamic crack front deformations. The process zone size at rest $\wo$ is kept relatively small compared to the heterogeneities size, and the contrast in toughness is achieved by varying the peak strength while keeping the process zone size at rest constant across the interface. According to \cite{morrissey_perturbative_2000}, a front dynamically stiffens with increasing propagation velocity and thus diminishes its deformations. We show in Fig.~\ref{fig:fig2}a the amplitude $A$ of front deformations as a function of the propagation velocity with $\Vc / \Cr \in [0.3-0.9]$ (black diamonds), with $\Cr$ the Rayleigh wave speed. The amplitude indeed decreases for faster cracks. The effect of dynamic stiffening on front deformation can be quantified by the function $D_I(\Vc)$ which only depends on the propagation velocity and whose derivation is given in \cite{supplemental}. The dashed black line in Fig.~\ref{fig:fig2}a is $D_I(\Vc) A_{\mathrm{lefm}}/d$, with $A_{\mathrm{lefm}}$ the predicted amplitude of front deformation based on the classical line tension model which is valid for small process zone size, and this function matches the amplitude observed in the simulations. Fig.~\ref{fig:fig2}b-d are snapshots of the crack front configuration for $\Vc / \Cr = 0.3 , 0.6 , 0.9$. The crack is shown in brown, the process zone size in orange, and the shades of green stand for the toughness of the intact part of the interface. In these snapshots, two effects of an increasing crack velocity are visible: (i) a decrease of the deformations and (ii) a decrease of the process zone size. The latter is known as the Lorentz contraction \cite{rice_mechanics_1980} of the process zone and is highly relevant for the following when we assess the effect of this length-scale on front deformation. The instantaneous process zone size for a mode I crack is given by $\wv  = \wo / A_I(\Vc)$ with $A_I$ a universal function of the crack velocity \cite{Freund1998}.

%%%%%%%%%%%%%% Figure %%%%%%%%%%%%%%%%%%%
\begin{figure*}[ht!]
%\centering
\includegraphics[width=1\textwidth]{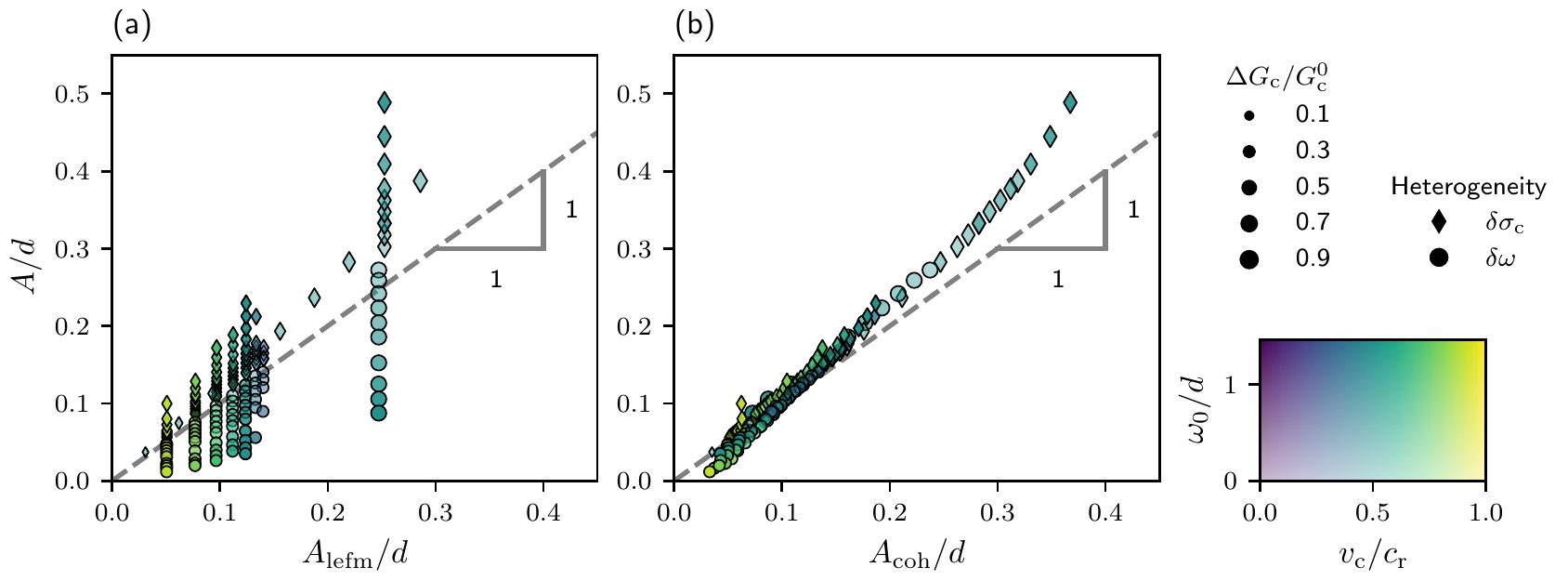}
\caption{The amplitude of the front deformation measured in simulations is compared with: (a) the prediction from the LEFM theory that does not account for a finite process zone, (b) the prediction from our newly derived dynamic cohesive "line tension" model. The simulations shown here explore a broad range of parameters, including variations of the crack propagation velocity $\Vc$, the process zone size at rest $\wo$ the toughness contrast $\Delta \Gc$ and the type of heterogeneities (constant process zone or constant peak strength). The detailed description of each data point is available in \cite{supplemental}.}
\label{fig:fig4}
\end{figure*}
%%%%%%%%%%%%%%%%%%%%%%%%%%%%%%%%%%%%%%%%%%%%%%%

\paragraph*{Process zone size and type of heterogeneities}
\label{sec:omega}

The influence of the process zone size is investigated. We consider two different cases: heterogeneities of peak strength $\sigmac$ (with constant process zone, see Fig~\ref{fig:fig1}d), and heterogeneities of process zone size at rest $\wo$ with constant peak strength, see Fig~\ref{fig:fig1}e). We vary in both cases the average value $\wo$ of the quasi-static process zone size while keeping the toughness contrast and the propagation velocity constant. The amplitude of crack front deformations is shown in Fig.~\ref{fig:fig3}a, for $\Vc = 0.5 \Cr$, $\Delta \Gc = 0.4 \Gc^0$ and $\wv / d \in [0.05 - 1.5]$ for both heterogeneities of peak strength (diamonds) and process zone size (circles). For small relative process zone size $\wv / d$ the front deformations amplitude is similar for both types of heterogeneities. However, they get significantly farther apart with increasing process zone size. On one hand, the amplitude increases with the dissipation length-scale for heterogeneities of peak strength (diamonds in Fig.~\ref{fig:fig3}a and snapshots in Fig.~\ref{fig:fig3}b-d). On the other hand, the amplitude diminishes with the process zone size for heterogeneities of process zone (circles in Fig.~\ref{fig:fig3}a and snapshots in Fig.~\ref{fig:fig3}e-f). Changes in process zone size are accommodated more easily by a crack front than changes in peak strength. These observations are striking: the deformations of a cohesive crack propagating through a heterogeneous microstructure is strongly dominated by the nature of the heterogeneities. For two interfaces sharing the same fracture toughness contrast, the difference between the two types of heterogeneities investigated in this work reaches up to a factor 4 when the process zone and the heterogeneities have the same size $\wv / d \sim 1$. The deformations are not tied directly to the toughness contrast, but rather to the variations of the cohesive parameters. For the slip weakening law used in this manuscript and heterogeneities that are achieved by varying both the peak strength and the process zone size (not presented in this manuscript), we expect the behavior to be bounded by the two limiting cases that were investigated. Note that this difference is expected to vanish for negligibly small relative process zone size, which can occur either with brittle materials or when cracks propagate at a velocity close to the limiting wave speed due to the Lorentz contraction.

\paragraph*{Theoretical Model}

In order to understand these surprising observations, we go back to the \emph{quasi-static cohesive line tension model} that has been recently derived in \cite{lebihain_quasi-static_2022}. This model extends \cite{rice_mechanics_1980} first-order theory by including the effect of cohesive stresses that resist the crack opening and is based on the weight functions associated to a point force located at a given distance from the front (i.e. inside the process zone). Two competing mechanisms arise from the presence of a cohesive zone : (i) the front \emph{stiffness} is reduced and (ii) the fluctuations of strength $\delta \sigmac$ and process zone $\delta \omega$ are smoothed out. In \cite{lebihain_quasi-static_2022}, it is predicted that these competing effects can have two different outcomes in the quasi-static regime: for heterogeneities of strength only, the front deformation amplitude is enhanced while for heterogeneities of process zone they are diminished. This is in qualitative agreement with the results reported in Fig.~\ref{fig:fig3}a. However, our simulations correspond to fully dynamic rupture while \cite{lebihain_quasi-static_2022}'s model is limited to quasi-static cracks. Two additional effects are expected to emerge when extending this model to dynamics: (iii) the process zone size changes dynamically with the propagation velocity: it shrinks when a crack accelerates due to the Lorentz contraction \cite{rice_mechanics_1980} and (iv) the front stiffens with increasing crack velocity \cite{morrissey_perturbative_2000}. For the same interface layout, a faster crack is expected to deform less, and the differences between the type of heterogeneities should be reduced. In order to validate our observations, we thus extend the \emph{quasi-static cohesive line tension model} of \cite{lebihain_quasi-static_2022} to dynamics in the permanent regime (i.e. constant  propagation velocity, see details in \cite{supplemental}) and obtain for the front deformations $\delta a$ :

\begin{equation}
\label{eq:FrontDeformation_Complete}
\widehat{\delta a}(k) = - D_I(\Vc) \left( \wv \dfrac{\hat{\Sigma}(\kw_v)}{\hat{\mathcal{A}}(\kw_v)} \dfrac{\widehat{\delta \sigmac}(k)}{\sigmac^0} + \wv \dfrac{\hat{\Omega}(\kw_v)}{\hat{\mathcal{A}}(\kw_v)} \dfrac{\widehat{\delta \omega}(k)}{2\wv} \right)
\end{equation}

with $k$ the wavenumber and $\hat{.}$ indicates a Fourier transform. $\wv$ is the instantaneous process zone size (related to (iii) above) and $D_I(\Vc)$ is a function of the velocity and represents the dynamic stiffening of the front (point (iv) above). $\hat{\mathcal{A}}$ and $\hat{\Sigma}$ and $\hat{\Omega}$ are functions of the nature of the weakening, the wavenumber $k$ and the process zone size. The exact formulation for these functions is given in \cite{supplemental}. The term $\hat{\mathcal{A}}$ acts as the \emph{loss of stiffness} of the front due to the introduction of a finite-size region of dissipation mentioned in point (i), while $\hat{\Sigma}$ and $\hat{\Omega}$ smooth out the fluctuations of material properties mentioned in point (ii). 

\paragraph*{Comparison between theory and simulations}

Crack front deformation simulations have been conducted for a broad range of parameters, including variations of process zone size at rest $\wo$, toughness contrast, type of heterogeneities, and crack front velocity $\Vc$. In Fig.~\ref{fig:fig4}a the front deformation amplitude measured from the simulations is plotted versus the prediction from the standard line tension model, including the dynamic stiffening term (from Eq.~\eqref{eq:FrontDeformation_Lefm}) that does not consider the existence of a finite dissipation length-scale near the crack tip. This prediction fails, as we have established previously that a finite process zone size strongly impacts the crack front deformations. For a given prediction based on the LEFM theory (take for example $A_{\mathrm{lefm}}/d = 0.25$) there is a large spread of measured amplitude, which can be either larger or lower than the predicted one (the dashed-gray line has a slope of 1) depending on the type of heterogeneities. It is expected from the observations of Fig.~\ref{fig:fig3} that simulations with a small process zone (e.g., for fast ruptures) will result in a significantly smaller difference between the two types of heterogeneities. This is apparent with the data points corresponding to fast cracks (yellow-green in Fig.~\ref{fig:fig4}a) that are significantly closer than the ones for slower cracks (blue data points). The effect of the front stiffening with increasing velocity is also visible from Fig.~\ref{fig:fig4}a, with large velocities resulting in small amplitudes.
In Fig.~\ref{fig:fig4}b, the prediction of Eq.\eqref{eq:FrontDeformation_Complete}, the \emph{dynamic cohesive line tension model}, is tested: all the data are falling close to a linear master curve, strongly supporting the validity of our model for rationalizing the effect of a finite process zone. While the predictions of Eq.~\eqref{eq:FrontDeformation_Complete} are based on the assumption of a semi-infinite crack, finite-size cracks have been considered in the simulations. Plus, the simulated ruptures are not in a steady permanent regime as assumed in the model. Second-order effects might also be required to accurately describe the deformations of cohesive fronts, as the latter can display larger curvatures than the classical line tension fronts.  This could potentially explain the small deviation from the predictions. Nonetheless, the proposed model successfully predicts the numerical observations and thereby the non-trivial influence of a finite dissipation length-scale for crack front deformations at constant propagation velocity: not only does the process zone influence front deformations, but also its outcome varies strongly depending on the description of the heterogeneities.

%TC:break Summary
\paragraph*{Discussion}

The deformations of a dynamic cohesive crack propagating through a heterogeneous field of toughness have been investigated numerically using the spectral boundary integral method coupled with a cohesive zone model. While the influence of the toughness contrast on front deformations amplitude is in agreement with the prediction of the classical line tension model (i.e. a linear increase of amplitude with contrast), modifying the process zone size introduces scale effects in the deformation of the crack front that are non-trivial. For the same toughness contrast and average process zone size, the crack front deformation amplitude is enhanced when considering heterogeneities of peak strength and diminished for heterogeneities of process zone. When considering the dynamics of the front, these differences are mitigated by the Lorentz contraction of the process zone size, and the amplitude of front deformations is decreased due to the dynamic stiffening of the front with increasing crack velocity. To rationalize these observations, we extended the \emph{cohesive line tension model} recently proposed in \cite{lebihain_quasi-static_2022} to dynamic rupture. This model predicts accurately the amplitude of the observed deformations, taking into account the instantaneous average process zone size and the propagation velocity. All in all, our model reveals the non-trivial effect of a finite dissipation length scale on the front deformations, and particularly the importance of the nature of the heterogeneities. Building a complete cohesive model including change in velocity and variations of properties along the front propagation direction remains a challenge. For the latter, the process zone size is expected to be also the relevant length scale, as the properties are averaged over the process zone size \cite{barras_interplay_2017}. This work provides the necessary ingredients to characterize the front roughness of disordered materials \cite{daguier_pinning_1997,delaplace_high_1999}, giving access to an estimate of the Larkin length. This directly impacts the prediction of the effective propagation threshold in cohesive composites\cite{demery_microstructural_2014}. Finally, this work might help understand the occurrence of out-of-plane damage as a consequence of high in-plane curvature of the front \cite{kolvin_nonlinear_2017}, and more generally the deformations of a three-dimensional crack front for which the process zone size changes with the orientation from the crack tip.

\bibliographystyle{apsrev4-2} 
\bibliography{PRL_FrontDeformation}

%% Supplemental material
\newpage
\pagebreak
\appendix
\setcounter{figure}{0}
\setcounter{equation}{0}
\renewcommand{\thefigure}{S\arabic{figure}}
\renewcommand{\theequation}{S\arabic{equation}}

\section{\Large{Supplemental Material}}

\subsection{Spectral Boundary Integral Method}
\label{app:BIM}

The simulations are performed using an in-house open-source implementation (called cRacklet~\cite{roch_cracklet_2022}) of the spectral boundary integral formulation of the elastodynamic equations~\cite{Geubelle1995,Morrissey1997,Breitenfeld1998}. This method describes the behavior at the interface between two semi-infinite elastic solids. The basic relation between the interfacial stresses $\bm{\sigma}$ and the opening displacements $\bm{u}^{\pm}$ in this case is given in Eq.~\eqref{eq:BIM}:

\begin{equation}
\label{eq:BIM}
\bm{\sigma}(\bm{x},t)^{\pm} = \bm{\sigma_{\infty}}^{\pm}(t) - \underline{\bm{V}}\cfrac{\partial \bm{u}^{\pm}}{\partial t}+\bm{s}^{\pm}(\bm{x},t)
\end{equation}

The $+$ and $-$ superscripts stand for the top and bottom solid. The first contribution is the remotely applied loading $\bm{\sigma_{\infty}}(t)$, the second is the so-called radiated damping term where $\underline{\bm{V}}$ is a diagonal matrix with

\begin{equation}
\label{eq:V}
    V_{xx} = V_{zz} = \mu / \Cs , V_{yy} = \mu \Cd / \Cs^2  
\end{equation}

$\mu$ the shear modulus, $\Cs$ and $\Cd$ respectively the shear wave speed and the longitudinal wave speed. $\bm{s}(x,t)$ represents the spatio-temporal interaction of different points on the interface mediated by bulk elastodynamics and is related to the interfacial displacement history through a convolution integral. Its Fourier representation can be found in~\cite{Breitenfeld1998}. Eq.~\eqref{eq:BIM} is completed by interface conditions (Eq.~\eqref{eq:CohesiveLaw}): as long as the stress at the interface is lower than the interfacial strength, continuity of tractions and displacements are satisfied at the interface. Otherwise, the interface is opening: the velocity is computed such that the stresses are in equilibrium with the strength of the interface given by Eq.~\eqref{eq:CohesiveLaw} as a function of the displacement jump. The displacement $\bm{u}(\bm{x},t)$ is then integrated in time using an explicit time-stepping scheme:

\begin{equation}
\bm{u}(\bm{x},t+\Delta t)\= \bm{u}(x,t) + \cfrac{1}{2} \cfrac{\partial \bm{u}(x,t)}{\partial t} \Delta{t}    
\end{equation}

with the time step being $\Delta{t}\= \alpha \Delta{x}/ \Cs$, where $\Delta{x}$ is the numerical grid spacing. The numerical parameter $\alpha$ is chosen to ensure the stability and the convergence of the numerical scheme, and is typically set to $0.2$. 
In our numerical simulations, the interface is initially at rest under homogeneous tensile stresses. A crack is slowly grown until it spontaneously propagates at the targeted velocity. The loading is tailored from a reference simulation in a 2D setup with homogeneous interfacial properties such that the crack velocity is constant during propagation.

\subsection{Material properties}

The simulations reported in the manuscript have been conducted using the elastic material properties of Homalite: Young's Modulus $E = 5.3e9$ [Pa], Poisson's ratio $\nu = 0.35$ [-] and shear wave speed $\Cs = 1263$ [m/s]. For the interface behavior, the fracture toughness $G_c^0 = 90$ [J/m\textsuperscript{2}] is defined by a couple of maximum stress and critical opening values between $(\sigma_c^0,\delta_c^0) = (7.79 \times $ 10\textsuperscript{6},$2.31 \times $ 10\textsuperscript{-5}$)$ [Pa.m] and $(\sigma_c^0,\delta_c^0) = (2.08 \times $ 10\textsuperscript{6},$8.64 \times $ 10\textsuperscript{-5}$)$ [Pa.m]
. The process zone at rest associated to these parameters goes from $\wo = 6.54\times $ 10\textsuperscript{-3} [m] to $\wo = 9.15\times $ 10\textsuperscript{-2} [m]. The full details of each simulation and the code used to run the simulations is available at \cite{zenodo}.

\subsection{Time evolution of the crack deformations and crack front waves}

\begin{figure}[ht!]
\centering
\includegraphics[width=0.5\textwidth]{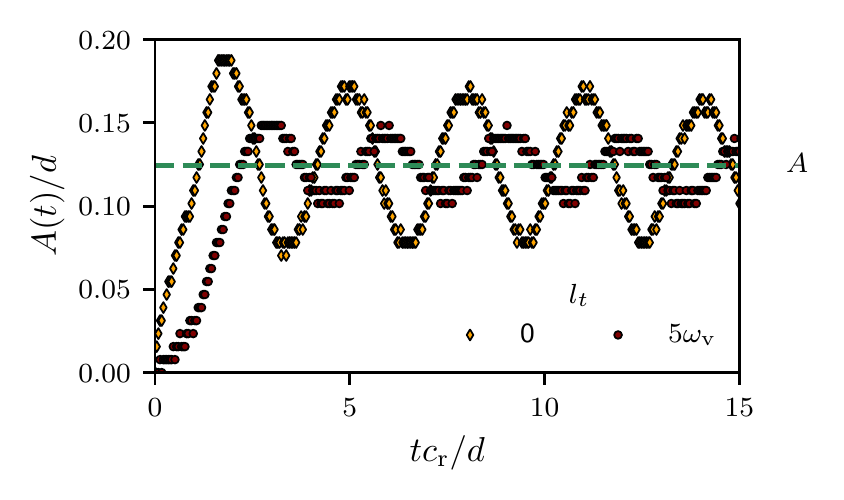}
\caption{Evolution of the amplitude of the front deformations as a function of time for two simulations with $\Vc = 0.7 \Cr$, $\Delta G_c / G_c^0 = 0.4$ and $\wo / d \simeq 0.42$. The yellow diamonds correspond to an interface with an abrupt change of properties $l_{\mathrm{t}} = 0$. The results corresponding to an interface with a gradual change of properties over the transition length $l_{\mathrm{t}} \simeq 5 \wv$ are shown with brown circles.  The green line indicates the steady-state amplitude around which the instantaneous amplitude oscillates.}
\label{fig:figS1}
\end{figure}

The deformation of the crack front is not instantaneous. When a crack starts interacting with a heterogeneous field of toughness, the perturbation propagates along the front via crack front waves \cite{morrissey_crack_1998,fekak_crack_2020}. If the change in toughness is abrupt, the front deformation amplitude overshoots its final value and then oscillates around it. The amplitude of these oscillations decreases slowly with time $\propto 1/\sqrt{t}$. As we are interested in the value of the equilibrium amplitude, we change progressively the toughness properties along a length $l_t$ to reduce the amplitude of these oscillations, such that the simulated cracks are closer to a permanent regime. We illustrate in Fig.~\ref{fig:figS1} the time evolution of the amplitude of the crack front normalized by the heterogeneity size in two cases: one with an abrupt change of toughness, i.e. $l_{\mathrm{t}} = 0$ (yellow diamonds) and a case with $l_{\mathrm{t}} = 5 \wv$ (brown circles). For these two simulations, $\Delta G_c / G_c^0 = 0.4$, $v_c / c_r = 0.7$ and $\wo / d \sim 0.42$. The oscillations of the front amplitude are significantly reduced when the material properties are slowly changed over the transition length $l_{\mathrm{t}}$. A longer transition length would diminish the oscillations even more, but would require to enlarge the length of the system and increase the computational cost.

%%%%%%%%%%%%%% Figure %%%%%%%%%%%%%%%%%%%
%%%%%%%%%%%%%%%%%%%%%%%%%%%%%%%%%%%%%%%%%%%%%%%

\begin{figure}[ht!]
\centering
\includegraphics[width=0.5\textwidth]{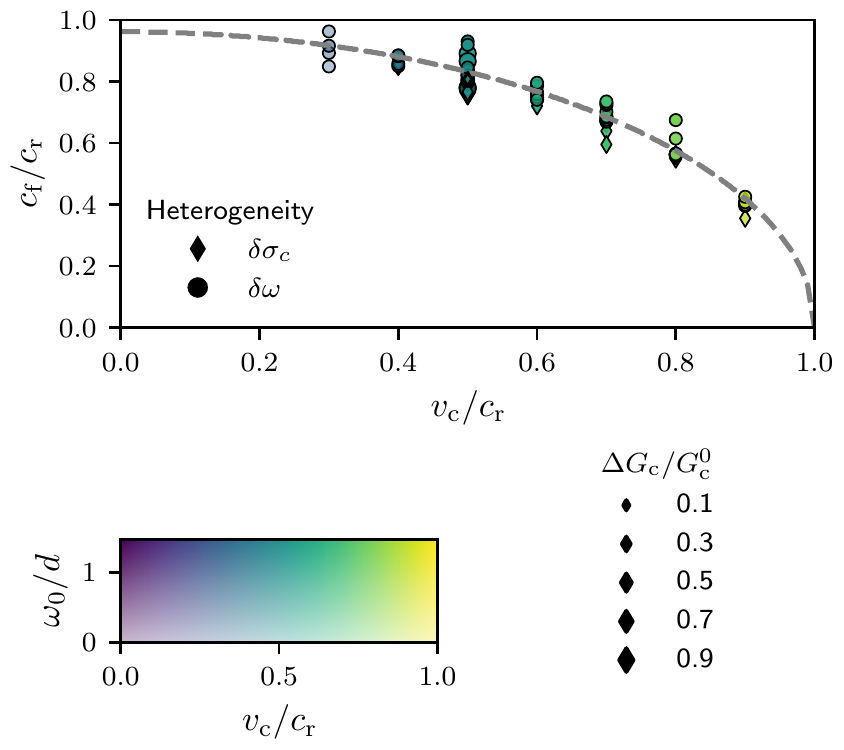}
\caption{Crack front wave velocity $\Cf$ as a function of the front velocity $\Vc$. The dashed gray line is the theoretical prediction following \cite{ramanathan_dynamics_1997}. Not all the simulations are shown in this figure as it is not always possible to determine the crack front wave velocity properly, see for example the case in Fig.~\ref{fig:figS1}a. }
\label{fig:figS2}
\end{figure}

The period of oscillations is characteristic of propagation velocity of the crack front waves: the time interval between two local extrema corresponds to the time that is required for the crack front waves to propagate across a distance $d$. When possible, we computed the velocity of the crack front waves and reported them in Fig.~\ref{fig:figS2}. Note that in some cases the oscillations are almost completely eliminated, and thus it is not possible to easily measure the velocity of the crack front waves. This is mostly the case for simulations with large process zone size. The change of properties in the $x$ direction is averaged over the process zone size, leading to an apparent change in toughness that is smoother and resulting in crack front waves with lower amplitude. The velocities of the crack front wave in our simulations are in agreement with the theoretical prediction given by \cite{ramanathan_dynamics_1997} (in dashed gray in Fig~\ref{fig:figS2}). The spread around the theoretical prediction for a given crack velocity is related to the difficulty in computing the crack front wave velocity. Contrarily to the case originally explored by \cite{morrissey_crack_1998} in mode I or later by \cite{fekak_crack_2020} in mode II, in which a single asperity creates a perturbation whose propagation along the front is clearly visible, the heterogeneous pattern investigated in this manuscript results in the front shape changing at every position along $z$ at the same time, leading to a challenging identification of the front wave velocity. The latter is computed as explained previously by identifying the period of oscillations, and thus requires finding local extrema of a discrete set of points. The procedure used here involves smoothing the data, which might alter slightly the precision of the results.

%%%%%%%%%%%%%% Figure %%%%%%%%%%%%%%%%%%%
%%%%%%%%%%%%%%%%%%%%%%%%%%%%%%%%%%%%%%%%%%%%%%%

\subsection*{Toughness contrast with constant process zone size}

\begin{figure}[ht!]
%\centering
\includegraphics[width=0.5\textwidth]{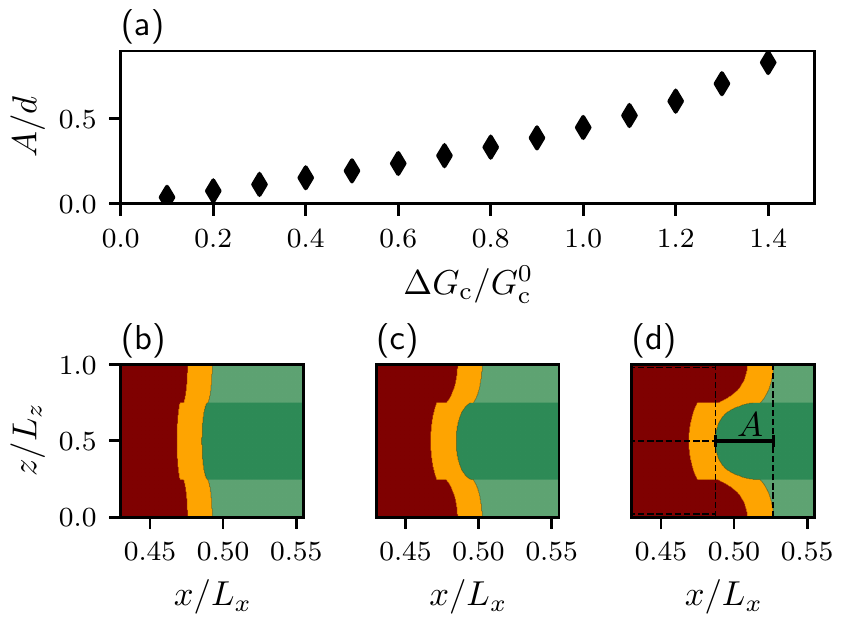}
\caption{(a) Scaling of the amplitude $A$ of the front deformations with the toughness contrast $\Delta \Gc / \Gc^0$. (b)-(c)-(d): snapshots of the crack front deformation for respectively $\Delta \Gc / \Gc^0 = 0.3 , 0.7 , 1.2$. The crack is in brown, the process zone in orange, the strong toughness in dark green and the weak one in light green. The crack velocity in these simulations is $\Vc = 0.5 \Cr$.}
\label{fig:figS3}
\end{figure}

To assess the validity of the numerical model, we first investigate the effect of the toughness contrast on the dynamic crack front deformations. The process zone is kept relatively small compared to the heterogeneities size, and the contrast in toughness is achieved by varying the peak strength while keeping the process zone size constant across the interface. The amplitude of the front deformations, normalized by the heterogeneities size, is shown in Fig.~\ref{fig:figS3}a as a function of the toughness contrast for $\Delta G_c / G_c^0 \in [0.1,1.4]$. Fig.~\ref{fig:figS3}b-d are snapshots of the crack front configuration for $\Delta G_c / G_c^0 = 0.3 , 0.7 , 1.2$. The crack is shown in brown, the process zone size in orange, and the shades of green stand for the toughness of the intact part of the interface. We observed a roughly linear increase of the front deformations with increasing fracture toughness contrast. For brittle materials (i.e. no process zone size), the Fourier transform of the quasi-static front deformations $\delta a$ is given by, see \cite{lebihain_quasi-static_2022}, 

\begin{equation}
\label{eq:FrontDeformation_Lefm}
\widehat{\delta a}(k) = - \dfrac{1}{|k|} \dfrac{\widehat{\delta \Gc(k)}}{\Gc^0}
\end{equation}

with $k$ the wavenumber and $\hat{.}$ indicates a Fourier transform. Eq.~\eqref{eq:FrontDeformation_Lefm} predicts a linear dependency of the front amplitude on the toughness contrast, which is consistent with our observations. For large contrasts, the observations deviate from the predictions, which is expected as second-order effects start being relevant.

%%%%%%%%%%%%%% Figure %%%%%%%%%%%%%%%%%%%
%%%%%%%%%%%%%%%%%%%%%%%%%%%%%%%%%%%%%%%%%%%%%%%

%\subsection{Effective toughness in the average along Oz (derive it again?)}

\subsection{Effect of dynamics on crack front deformations}
\label{seq:derivationDynamic}

\begin{figure}[ht!]
%\centering
\includegraphics[width=0.5\textwidth]{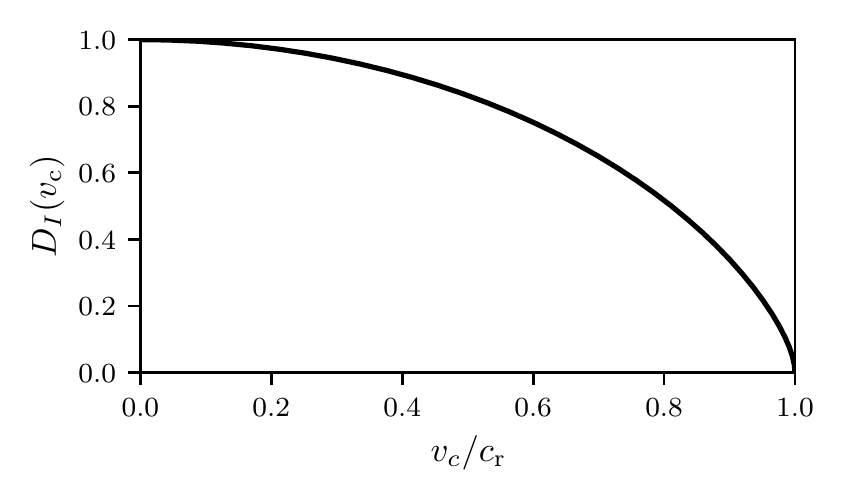}
\caption{Dynamic pre-factor $D_I(\Vc)$ from Eq.~\eqref{eq:Dv} as a function of the propagation velocity $v_c/c_r$.}
\label{fig:figS4}
\end{figure}

We now wish to assess the influence of dynamics on the crack front deformations. We will consider a permanent regime, i.e. a crack that has been propagating at a constant velocity for an infinite amount of time. An expression for the perturbation of the dynamic stress intensity factor for a small deviation from straightness of a crack is provided by \cite{willis_dynamic_1995}. The general structure of the equation relates the perturbed stress intensity factor to the original stress intensity factor and a convolution of the front deformation with a function $P$, see Eq.~(8.10) in \cite{willis_dynamic_1995}. For a mode I crack, its perturbed stress intensity factor $K_I$ writes as Eq.~\eqref{eq:PerturbedK}. $K_I^0$ is the stress intensity factor in the unperturbed configuration, and $\mathrm{PV}$ denotes a Cauchy principal value. $P(z,t)$ is a kernel whose expression in the wavenumber-frequency domain $(z \to k),(t \to \theta)$ is given by \cite{ramanathan_dynamics_1997,morrissey_perturbative_2000}. We consider only the permanent regime for which there is no time dependency ($\theta = 0$) and in this case $P(z,t)$ reduces to $D_I(v) |k| / 2$ with $D_I(v)$ given by Eq.~\eqref{eq:Dv}. It corresponds to the dynamic stiffening term associated with mode I solicitation. We show the function $D_I(v)$ in Fig.~\ref{fig:figS4}. It tends towards 1 for the quasi-static case $\Vc = 0$ and towards 0 for cracks approaching the limiting propagation velocity, the Rayleigh wave speed $\Cr$.

\begin{widetext}
\begin{equation}
    K_I(z,t) = K_I^0(z) + \delta K_I(z,t) = K_I^0(z) + \dfrac{\delta K_I^0}{\partial a}(z) \delta a(z,t) - \mathrm{PV} \int_{-\infty}^{+\infty} P(z,t) K_I^0(z') \left[ \delta a(z,t)-\delta a(z',t) \right] dz'
\label{eq:PerturbedK}
\end{equation}

\begin{equation}
\label{eq:Dv}
D_I(\Vc) = 1 / \left( \dfrac{2}{\sqrt{1-(\Vc/\Cr)^2}} - \dfrac{1}{\sqrt{1-(\Vc/\Cd)^2}} \\ - (\Vc/\Cr)^2 \int_{\Cs}^{\Cd} \phi(v) dv \right)
\end{equation}

\end{widetext}

\subsection{Dynamic Cohesive Line Tension Model}
\label{seq:derivationDynamicLineTension}

In order to derive a dynamic cohesive line tension model, one can build on the derivation for the quasi-static cohesive line tension model of Lebihain et al. \cite{lebihain_quasi-static_2022} to compute the expression of the stress intensity factor $k$ of the deformed front $\mathcal{F}^*$ that is generated at a point $z=z_0$ by a pair of unitary forces that are applied at a given distance $x$ behind the crack front at a point $z=z_1$, see Eq.~(7) in \cite{lebihain_quasi-static_2022}. In the permanent dynamic regime, it writes

\begin{widetext}
\begin{equation}
\label{eq:DynamicPerturbedWeightFunction}
k\left(\mathcal{F}^*; z_0, z_1, x,\Vc\right) = k\left(\mathcal{F}; z_0, z_1,x,\Vc\right) + D_I(\Vc) \int_{-\infty}^{+\infty} k\left(\mathcal{F}; z; z_1, x,\Vc\right) \dfrac{\delta a(z)-\delta a(z_1)}{(z-z_1)^2} dz
\end{equation}
\end{widetext}

where $k\left(\mathcal{F}; z_0, z_1, x,\Vc=0\right)$ is known analytically for the semi-infinite coplanar crack with a straight crack front $\mathcal{F}$, see \cite{lebihain_quasi-static_2022} for more details on the derivation of the crack face weight functions.

The derivation of the dynamic cohesive line tension model follows then the one presented in \cite{lebihain_quasi-static_2022} for the crack front waves weight functions and the cohesive stress intensity factor, with the difference that the process zone to be considered is the instantaneous cohesive zone size $\wv$ instead of the rest one $\wo$, and the pre-factor $D_I(\Vc)$ multiplying the terms. The complete prediction for the deformation of a front in the dynamic regime due to both heterogeneities of strength and process zone thus corresponds to Eq.~(53) in \cite{lebihain_quasi-static_2022} with the two changes mentioned above, which result in Eq.~\eqref{eq:FrontDeformation_Complete}.
Note that as $D_I(\Vc = 0) = 1$, we recover the formulae given by \cite{lebihain_quasi-static_2022} for the quasi-static front deformation in presence of a process zone. For cracks propagating at the limiting velocity, we have $D_I(\Vc = \Cr) = 0$, resulting in theoretically undeformable crack front in this limit (in the hypothesis of co-planar crack propagation). In practice, fast cracks will often trigger out-of-plane damage and instabilities before reaching the limiting velocity.

%%%%%%%%%%%%%% Figure %%%%%%%%%%%%%%%%%%%
%%%%%%%%%%%%%%%%%%%%%%%%%%%%%%%%%%%%%%%%%%%%%%%

For completeness, we recall here the expression for $\hat{\mathcal{A}}$ and $\hat{\Sigma}$ and $\hat{\Omega}$. Note that these expressions slightly differ from the one given by \cite{lebihain_quasi-static_2022} as we consider here the dynamic process zone size $\wv$ and not the static one $\wo$.

\begin{equation}
\label{eq:Fourier_Cohesive_Prefactors}
\begin{cases}
\hat{\mathcal{A}}\left(\kw_v\right) & = -\dfrac{1}{\Cw} \int_{0}^{+\infty} \dfrac{\dfw\left(u\right)}{u^{1/2}} \left(1-e^{-\kw_v u}\right) du \\
\hat{\Sigma}\left(\kw_v\right) & = \dfrac{1}{\Cw} \int_{0}^{+\infty} \dfrac{\fw\left(u\right)}{u^{1/2}} e^{-\kw_v u} du \\
\hat{\Omega}\left(\kw_v\right) & = -\dfrac{2}{\Cw} \int_{0}^{+\infty} \dfw\left(u\right) u^{1/2} e^{-\kw_v u} du
\end{cases}
\end{equation}

with $\Cw = \int_{0}^{+\infty} \fw\left(u\right) u^{-1/2} du$ and $\fw(x/\omega)$ the shape function that relates to the nature of the weakening. For the linear traction separation law considered in this work there is no analytical expression for the shape function as a function of the distance, but it can be computed numerically, see \cite{lebihain_quasi-static_2022} Appendix C.4. for details.

\end{document}